\documentclass[preprint,aps,onecolumn,nofootinbib]{revtex4}
\usepackage{dcolumn}
\usepackage{bm}
\usepackage{mathrsfs}
\usepackage{amssymb}
\usepackage{amsmath}
\usepackage{amsfonts}
\usepackage{color}
\usepackage{ulem}
\begin{document}
\bibliographystyle{prsty}
\title{Local causality in the works of Einstein, Bohm and Bell\footnote{Accepted for publication in A. Oldofredi (Ed.), Guiding Waves in Quantum Mechanics: 100 Years of de Broglie-Bohm Pilot-Wave Theory, Oxford University Press.}}
\author{ Aur\'elien Drezet$^{1}$}
\address{(1) Institut NEEL, CNRS and Universit\'e Grenoble Alpes, F-38000 Grenoble, France }
\email{aurelien.drezet@neel.cnrs.fr}
\begin{abstract}
 In this chapter we discuss the Einstein Podolsky Rosen theorem and its strong relation with Bell's theorem. The central role played by the concept of beable introduced by Bell is emphasized. In particular we stress that  beables involved in EPR and Bell theorems are not limited to hidden supplementary variables (e.g., like in the de Broglie-Bohm (dBB) pilot-wave theory) but also include the wave function.  In full agreement with Bell this allows us the reformulate the EPR and Bell results as strong theorems concerning nonlocality for quantum mechanics itself and not only for  hidden-variables approaches as it is often mistakenly assumed. Furthermore, we clarify some  repeated ambiguities concerning `local-realism' and emphasize that neither realism nor determinism nor counterfactual definiteness are  prerequisites of EPR and Bell theorems.     
\end{abstract}

\maketitle
\section{Introduction and motivations}
\indent    David Bohm is a central figure in the history of quantum mechanics (QM). As it is well known in 1952 \cite{Bohm1952} he proposed  a deterministic hidden-variables theory able to complete standard QM and to reproduce its statistical predictions and results of standard QM. Moreover, the theory he developed was actually a rediscovery of the old pilot-wave theory  presented by Louis de Broglie in 1927 at the fifth Solvay conference.  Watched retrospectively seventy years later the most original contribution of Bohm compared to what de Broglie did  is perhaps his analysis of the Einstein Podolsky Rosen  (EPR) paradox concerning nonlocality and completeness of QM.\footnote{In this context we point out that de Broglie rejected nonlocality and criticized Bell's theorem \cite{Broglie1974}.} Indeed, the goal of the EPR article \cite{EPR} was to show that if we assume the principle of Einstein locality QM must be incomplete. More precisely EPR shows that either QM is incomplete or quantum  mechanics is nonlocal i.e., it violates Einstein's locality principle.  In this context, Bohm showed \cite{Bohm1952} that his own hidden-variables theory able to complete QM is explicitly non-local. While this doesn't contradict the EPR results seen as a theorem Bohm approach  was certainly the option `which Einstein would have liked least'. \footnote{For a discussion of the status of special relativity and nonlocality in relativistic dBB theories see the contribution of Valia Allori in this volume. } Of course this was just the beginning of the story: In 1964 John Bell, based on EPR work, discovered his famous theorem (\cite{Bell}, chap.~2)  firmly establishing  that QM (irrespectively of being complete or incomplete) must be nonlocal.\\
\indent The previous summary reflects the position advocated by some physicists  and   philosophers, including Sheldon Goldstein \cite{Bell2023}, Tim Maudlin \cite{Maudlin2014}, Jean Bricmont~\cite{Bricmont2016}, Travis Norsen~\cite{Norsen2006,Norsen2011},  David Albert \cite{Albert} and of course John Bell \cite{Bell}, but this is not the majority view. The majority view claims that Bell's theorem is concerned with `local realism' i.e., that it forces us to either abandon realism or locality. Since most physicists would disapprove relaxing locality they actually see the theorem as a strong indication that one must abandon realism. Most confusions concerning EPR, Bell's theorem and nonlocality arise from wavy and vague arguments associated with the definitions of Einstein locality and  realism.     In order to celebrate the work of Bohm on nonlocality  the present author thinks it could be useful to summarize once more the EPR-Bohm-Bell connection.
\section{The story} 
\indent Everything started with Einstein's disappointment about the way QM was built and axiomatized as a complete indeterministic theory. For Einstein the fact that quantum mechanics is a statistical theory should be explained by a deeper dynamical approach (probably deterministic) able to complete the statistical predictions by a classical-like type of mechanical and realistic explanations. One should not forget that Einstein was a pioneer of classical statistical mechanics with his famous 1905 interpretation of Brownian motion as resulting of an underlying mechanical process. He also developed general relativity which is a fundamentally deterministic theory of gravitation and space-time. Hence, Einstein could certainly not believe that God plays dice at the microscopic scale! Moreover, for him realism was a central prerequisite (more important even than determinism). In 1953 in a \textit{Festschrift} book to honor de Broglie achievements Einstein wrote: \textit{I am not blushing to put the concept of `real state of a system' at the center of my meditation} (\cite{Einstein1953}, p. 7).\\
\indent  In 1927 during the Solvay conference, he debated with Niels Bohr on the possibility to beat or defeat  the Heisenberg principle using correlations between a microscopic quantum particle (involved in a double-slit interference experiment) and some macroscopic  a priori classical devices entangled with the particle.  Bohr  recollection~\cite{Bohr1949} of these discussions shows that Einstein at that time underestimated the coherence of QM  and didn't realize that the entanglement between particles and macroscopic systems in his `which-path' experiment only makes sense if the macroscopic apparatus are also analyzed within QM.\footnote{Retrospectively, one can say that the most astonishing fact for Einstein  was perhaps the existence of Heisenberg's shift cut or split allowing one to describe any measuring apparatus as quantum or classical in order to remove paradoxes involving the uncertainty principle. Heisenberg also applied this approach in his unpublished response to EPR \cite{Crull}.}  In 1930 at the sixth Solvay conference, Einstein and Bohr continued this debate  with the famous `photon-box' Gedankenexperiment where Einstein attempted to circumvent the Heisenberg principle (in the time-energy domain) by using correlations between a particle in a box and a device measuring its time and energy (a detailed analysis of the paradox is provided in \cite{Nikolic}).  Once more, Bohr showed to Einstein that it is not allowed to contradict QM with this kind of approach since QM must be used for describing  every parts of the whole indivisible system in agreement with the complementarity principle. Moreover, these two encounters with Bohr convinced Einstein that QM is mathematically and physically self-consistent, and that it is illusory to try defeating Heisenberg's principle directly. However, he didn't stop the fight: in 1935 with his two collaborators Podolsky and Rosen he proposed the famous paradox involving entanglement between two remote particles \cite{EPR}. The core of the EPR article is the completeness of QM. By a complete  physical theory they meant that:  `every element of the physical reality must have a counterpart in the physical theory'. Questioning the completeness of QM is therefore suggesting that some effects or correlations having an empirical nature  are not predicted or explained by the theory. Of course, since QM is supposed to describe everything this would be impossible. Yet, EPR showed that by associating to QM a very natural and intuitive feature of the classical world (i.e., that nobody would like to abandon) we obtain a contradiction and therefore conclude that QM cannot be complete and at the same time satisfy this natural property.  This natural and intuitive property that EPR added is of course Einstein locality: a fundamental feature of causality in relativistic space-time.\\
\indent The whole EPR deduction is based on the  position and momentum observables for two entangled particles but it is common  (and we will follow this strategy) to use instead the example proposed by Bohm in 1951 of two spin-$\frac{1}{2}$ particles 1 and 2 entangled in the singlet state:
 \begin{eqnarray}
 |\psi\rangle=\frac{1}{\sqrt{2}}(|+1_{\mathbf{\hat{z}}}\rangle_1|-1_{\mathbf{\hat{z}}}\rangle_2-|-1_{\mathbf{\hat{z}}}\rangle_1|+1_{\mathbf{\hat{z}}}\rangle_2)
 \end{eqnarray} where $|+1_{\mathbf{\hat{z}}}\rangle_j,|-1_{\mathbf{\hat{z}}}\rangle_j$ are the two eigenstates of the $\sigma_z^{(j)}$ Pauli matrices for the $j^{th}$ particle. As a consequence we have $(\sigma_z^{(1)}+\sigma_z^{(2)})|\psi\rangle=0
$ that expresses the perfect anticorrelation between the two $z-$spin components. Importantly, by symmetry we have also $(\sigma_x^{(1)}+\sigma_x^{(2)})|\psi\rangle=0=(\sigma_y^{(1)}+\sigma_y^{(2)})|\psi\rangle$ expressing the perfect spin anticorrelation in every direction $x$, $y$ etc...    Due to this perfect anticorrelation between  the particles, experimentalists Alice and Bob recording respectively the projected spin of  particle 1 and 2 along a common axis  $\mathbf{\hat{n}}$ will naturally obtain the joint probability  
\begin{eqnarray}
P(\alpha,-\alpha/\mathbf{\hat{n}}_1,\mathbf{\hat{n}}_2=\mathbf{\hat{n}}_1,|\psi\rangle)=\frac{1}{2}, &
P(\alpha,\alpha/\mathbf{\hat{n}}_1,\mathbf{\hat{n}}_2=\mathbf{\hat{n}}_1,|\psi\rangle))=0
\end{eqnarray} where $\alpha=\pm 1$, and the probabilities are conditioned on the common analysis direction $\mathbf{\hat{n}}_1=\mathbf{\hat{n}}_2=\mathbf{\hat{n}}$ of the two independent Stern-Gerlach splitters used by Alice and Bob and by the singlet wave function $|\psi\rangle$. These conditions explicit the perfect anticorrelation at the core of EPR work. \\
\indent Now comes the crux: EPR introduce the Einstein-locality assumption based on natural features of the classical world picture concerning  correlations and relativistic causality. The main idea is that a local operation made by Alice on particle 1 at space-time point $x_1$ should not influence what is happening to the second particle recorded by Bob at space-time point $x_2$ if the two events are space-like separated (so that no-signal could propagate between the two points).  As Einstein wrote in 1949:
 \begin{quote}
 But on one supposition we should, in my opinion, absolutely hold fast: the real factual situation of the system $S_2$ is independent of what is done with the system $S_1$ which is spatially separated from the former (\cite{Einstein1949}, p. 85).
\end{quote}  
Assuming this natural relativistic  hypothesis the EPR deduction is easy to understand. From Eq. 2 we have perfect anticorrelation and therefore if Alice is recording the spin-1/2 component of particle 1 along  the direction $\mathbf{\hat{n}}_1=\mathbf{\hat{z}}$  with the result $+1$ (respectively $-1$) we know that Bob must necessarily record in his lab the result $-1$ (respectively) $+1$) for the same settings $\mathbf{\hat{n}}_2=\mathbf{\hat{z}}$. Moreover, nothing obliges Bob to measure the spin projection of his particle along the same direction and he could for instance decide to record the spin along the $x-$direction: $\mathbf{\hat{n}}_2=\mathbf{\hat{x}}$. Now Bob or Alice could take their decision at the last moment,  and since the locality principle is assumed to occur no influence that could modify their results are allowed to propagate. As a direct consequence  we can make a counterfactual reasoning:   if Bob records along the $x-$direction  and obtains for example $+1$ and Alice obtained for example the result $-1$ along the $z-$direction, then we know for sure that Alice should have obtained the result $-1$ along the  $x-$direction and Bob the result $+1$ along the $z-$direction, even thougth these experiments have not actually been done. Crucially the legitimacy of the counterfactual reasoning of EPR is mandated by Einstein-locality and is not an independent hypothesis.  This justifies the famous introduction of `elements of reality' by EPR: 
\begin{quote}
If, without in any way disturbing a system, we can predict with certainty (i.e., with probability equal to unity) the value of a physical  quantity, then there exists  an element of physical reality corresponding to this physical quantity.\cite{EPR}
\end{quote}   
Of course EPR understood perfectly well that counterfactual reasoning is in general forbidden for a single particle using the usual approach to QM. This is because   for a single particle one could invoke  Heisenberg's principle to impose a strong form of complementarity and contextuality: it is impossible to record in one single experiment the spin projection for the $x$ and $z$ direction because $\sigma_x$ and  $\sigma_z$  don't commute. Therefore, one must choose between one experiment or the other and if one is doing sequential experiments it is known that dispersion will occur in agreement with Heisenberg's principle. To speak about determinations and counterfactual hidden properties that could be observed but actually are not is considered as useless in standard QM. To cite Asher Peres: `Unperformed experiments have no results' \cite{Peres1978}.  However, employing Einstein-locality EPR found a clean way to go around Heisenberg's principle limitations.  Assuming locality we can know counterfactually the spin components of the two particles along orthogonal directions $x$ and $z$ even though we only actually measured the spin of particle 1 (respectively 2) along the $z$ (respectively $x$) direction. WSimilarly, with two measurements  and assuming locality we actually know four spin projection values along directions $x-z$ for the two particles, and this is impossible if we assume that QM is complete. In other words, assuming that  QM  is local (QM-L) and complete (QM-C) we conclude that QM is incomplete (QM-IC)!\\
\indent  This is a wonderful logical contradiction that can be formally written:
``QM-L $\& $ QM-C=False'' or equivalently  ``$\lnot$ QM-L OR  $\lnot$ QM-C=True'', i.e., ``QM-NL OR  QM-IC=True'' with  $\lnot$ QM-L is the negation of QM-L, i.e., QM is nonlocal and similarly $\lnot$ QM-C is actually QM-IC. EPR theorem is unavoidable,  it implies that if QM-L is true then we must have QM-IC true; and contrapositively if QM-C is true we must have  QM-NL true:
\begin{eqnarray}
\textrm{QM-L} \Rightarrow   \textrm{QM-IC},&\textrm{and }&
\textrm{QM-C}  \Rightarrow  \textrm{QM-NL}.
\end{eqnarray} 
Importantly,  EPR leads to three alternatives:
\begin{eqnarray}
(i) \textrm{QM-L}  & \& &  \textrm{QM-IC}\nonumber\\
(ii) \textrm{QM-NL}  & \& &  \textrm{QM-IC}\nonumber\\
(iii) \textrm{QM-NL}  & \& &  \textrm{QM-C}\label{EPR3}
\end{eqnarray} where (i) was favored by Einstein, (ii) by Bohm, and (iii) by Bohr and followers.\footnote{A different way to explain this is that either QM is incomplete  (i.e., regrouping options (i) and (ii) of Eq. \ref{EPR3}) \textit{or} (and this or is exclusive) QM is complete and  nonlocal, i.e., option (iii) of (\ref{EPR3}). } The beauty and logic of the EPR deduction/theorem is often underappreciated and the fact that counterfactuality and determinism are actually derived and not inferred by EPR is still nowadays misunderstood.   
\section{Bell's beables}
\indent An important comment must be done concerning EPR theorem. Indeed, while this result is rigorous, EPR didn't present it in a very formal way. The notions of locality and determinism were used in a very intuitive sense as shown, for instance, in the previous quote of Einstein concerning locality. Therefore one would have to be more precise on the definition of locality. This step was taken by Bell (as discussed below). Furthermore, the path of EPR to incompleteness implied determinism  meaning that we actually have:  QM-L $\Rightarrow$ QM-D i.e., QM is deterministic,  and from that   QM-D $\Rightarrow$ QM-IC.  But this derived notion of determinism is not very clearly discussed by EPR. For example, considering the actual $z-$component of the spin-1/2 particle measured by Alice  we see that the `element of reality' $v(\sigma_z^{(1)})$ associated with particle 1 implies through locality the existence of a couterfactual element of reality $v(\sigma_z^{(2)})=-v(\sigma_z^{(1)})$ for the second particle even if Bob actually measured    $v(\sigma_x^{(2)})$. Importantly, EPR don't require that  the observed values $v(\sigma_z^{(2)})$ preexist before the measurement. For example, in the dBB theory the spins are observables that are predetermined by the contextual dynamics. Therefore, in a deterministic and local theory we only assume that something before the measurement predetermined the elements of reality observed (or not) by Alice and Bob.   Writing  $\lambda$ this initial condition/predetermination we see that actually EPR determinism presupposed only that we have two functions 
$v(\sigma_n^{(1)}):=A(\lambda,\mathbf{\hat{n}},|\psi\rangle)$, $v(\sigma_n^{(2)}):=B(\lambda,\mathbf{\hat{n}},|\psi\rangle)=-A(\lambda,\mathbf{\hat{n}},|\psi\rangle)$ with $\sigma_n^{(i)}=\mathbf{n}\cdot\boldsymbol{\sigma}^{(i)}$ the spin operator of particle $i=1$ or 2 along the analysis direction $\mathbf{n}$. \\                
\indent In order to write the EPR reasoning in a formal way and discuss locality for deriving determinism, we must use Bell notations for the probabilities of beables. Bell introduced the concept of `beable' (\cite{Bell}, chaps.~5,7,16,24) to characterize every physical  (i.e., actual or real) properties belonging to the system that must influence the measured correlations. Crucially here  the beables must include the `classical' Stern-Gerlach devices with (in general different) directions $\mathbf{\hat{n}}_j$ (i.e., associated with external fields and potentials acting on the two parts of the quantum system under study), the possible hidden supplementary variables $\lambda$ (e.g., the particle positions in  the dBB pilot-wave theory), and the wave function $|\psi\rangle$.  It is unfortunate that most commentators of Bell (as an exception see \cite{Oldofredi}) fail to realize that $|\psi\rangle$ actually belongs to the fundamental beables listed by Bell.
Now it is always possible to write the quantum joint probabilities $P(\alpha,\beta/\mathbf{\hat{n}}_1,\mathbf{\hat{n}}_2,|\psi\rangle)=|\langle\alpha_{\mathbf{\hat{n}}_1},\beta_{\mathbf{\hat{n}}_2}|\psi\rangle|^2$ as:
\begin{eqnarray}
P(\alpha,\beta/\mathbf{\hat{n}}_1,\mathbf{\hat{n}}_2,|\psi\rangle)=\int_{\Lambda} P(\alpha,\beta/\lambda,\mathbf{\hat{n}}_1,\mathbf{\hat{n}}_2,|\psi\rangle)\rho(\lambda,\mathbf{\hat{n}}_1,\mathbf{\hat{n}}_2,|\psi\rangle)d\lambda
\label{Bellmodel1}
\end{eqnarray} 
where $\alpha,\beta=\pm 1$ and the integral spans over the hidden-variable space $\Lambda$.\footnote{We have $\int_{\Lambda} \rho(\lambda,\mathbf{\hat{n}}_1,\mathbf{\hat{n}}_2,|\psi\rangle)d\lambda=1$ and $\sum_{\alpha,\beta}P(\alpha,\beta/\lambda,\mathbf{\hat{n}}_1,\mathbf{\hat{n}}_2,|\psi\rangle)=1$. } At each run of the quantum experiment an actual value of $\lambda$ is selected. Bell used this notation in 1964 and 1971 (\cite{Bell}, chaps.~2,4) to derive  his theorem but actually we could also generalize it as 
\begin{eqnarray}
P(\alpha,\beta/\mathbf{\hat{n}}_1,\mathbf{\hat{n}}_2,|\psi\rangle)=\int_{\Omega} P(\alpha,\beta/\omega,\mathbf{\hat{n}}_1,\mathbf{\hat{n}}_2,|\psi\rangle)\rho(\omega,\mathbf{\hat{n}}_1,\mathbf{\hat{n}}_2,|\psi\rangle)d\omega\label{Bellmodel3}
\end{eqnarray} which is formally the same as the original formula Eq.~\ref{Bellmodel1} used by Bell in his 1964 paper. We introduced the beable $\omega:=(\lambda,\theta)$ where $\theta$ is a new beable describing the quantum state and $\Omega$ the full `ontic' space. Again a value of $\theta$ is actualized at each run of the experiment. Of course, since the probability description contains already $|\psi\rangle$ as a condition, we expect $\theta$ to be somehow redundant. For instance, according to Beltrametti and Bugajski\cite{Beltrametti1995,Drezet2012} for a complete theory (i.e., without $\lambda$) we can always write: 
\begin{eqnarray}
P(\alpha,\beta/\mathbf{\hat{n}}_1,\mathbf{\hat{n}}_2,|\psi\rangle)
=\int_{\Theta}  P(\alpha,\beta/\mathbf{\hat{n}}_1,\mathbf{\hat{n}}_2,|\theta\rangle)\delta(|\theta\rangle-|\psi\rangle)d|\theta\rangle\label{beltra}
\end{eqnarray}  where now the beable $|\theta\rangle$ belongs to the Hilbert space of the two-spins system equivalent to $\Theta$. We have $P(\alpha,\beta/\mathbf{\hat{n}}_1,\mathbf{\hat{n}}_2,|\theta\rangle)=|\langle\alpha_{\mathbf{\hat{n}}_1},\beta_{\mathbf{\hat{n}}_2}|\theta\rangle|^2$ and $\rho(\mathbf{\hat{n}}_1,\mathbf{\hat{n}}_2,|\psi\rangle,|\theta\rangle)=\delta(|\theta\rangle-|\psi\rangle)$.\footnote{More rigorously writing $|\theta\rangle=\sum_{\alpha,\beta}\theta_{\alpha,\beta}| \alpha_{\mathbf{\hat{n}}_1},\beta_{\mathbf{\hat{n}}_2}  \rangle$ with $\theta_{\alpha,\beta}=\theta'_{\alpha,\beta}+i\theta''_{\alpha,\beta}:=\langle\alpha_{\mathbf{\hat{n}}_1},\beta_{\mathbf{\hat{n}}_2}|\theta\rangle\in \mathbb{C}$ a generally complex valued amplitude (with $\alpha,\beta=\pm 1$) we have $\delta(|\theta\rangle-|\psi\rangle):=\prod_{\alpha,\beta}\delta(\theta'_{\alpha,\beta}-\psi'_{\alpha,\beta})\delta(\theta''_{\alpha,\beta}-\psi''_{\alpha,\beta})$ with $\psi_{\alpha,\beta}:=\langle\alpha_{\mathbf{\hat{n}}_1},\beta_{\mathbf{\hat{n}}_2}|\psi\rangle$,  and $d|\theta\rangle:=\prod_{\alpha,\beta}d\theta'_{\alpha,\beta}d\theta''_{\alpha,\beta}$. We have also $P(\alpha,\beta/\mathbf{\hat{n}}_1,\mathbf{\hat{n}}_2,|\theta\rangle)=|\theta_{\alpha,\beta}|^2$.} This representation corresponds to what Harrigan and Spekkens \cite{Harrigan2010} called  an ontic description of the quantum state where the Dirac distribution associates a strongly localized density of probability to $|\psi\rangle$. It is interesting to note that this is not the only ontic representation of the EPR state.  For example, inspired by Wigner phase distribution Scully \cite{Scully1983} developed an angular representation of the $|\psi\rangle$ state also involving a Dirac distribution. Scully specifically considered  the joint probability $P(\alpha,\beta/\mathbf{\hat{n}}_1,\mathbf{\hat{n}}_2,|\psi\rangle):=P(\alpha,\beta/\varphi_1,\varphi_2,|\psi\rangle)$ where $\mathbf{\hat{n}}_1,\mathbf{\hat{n}}_2$ are two apriori different unit vectors contained in the $x-z$ plane (the particles move along the two opposed $\pm y$ directions and $\varphi_j:=\widehat{\mathbf{\hat{z}},\mathbf{\hat{n}}_j}$ are the angles between the vector $\mathbf{\hat{n}}_j$ characterizing the spin analyzers  and the vertical  common axis $\mathbf{\hat{z}}$). He found:
\begin{eqnarray}
P(\alpha,\beta/\varphi_1,\varphi_2,|\Psi\rangle)=\iint P(\alpha/\theta_1,\varphi_1,|\Psi\rangle)P(\beta/\theta_2,\varphi_2,|\Psi\rangle)
\rho(\theta_1,\theta_2/\varphi_1,\varphi_2,|\Psi\rangle)d\theta_1 d\theta_2\label{scully}
\end{eqnarray} with $P(\alpha/\theta_i,\varphi_i,|\psi\rangle)=\frac{1+\alpha\cos{(\varphi_i-\theta_i)}}{2}$ and $\rho(\theta_1,\theta_2/\varphi_1,\varphi_2,|\Psi_{EPR})=\frac{1}{2}\delta(\theta_2-\theta_1-\pi)[\delta(\theta_1-\varphi_1)+\delta(\theta_1-\varphi_1-\pi)]$. This leads to $P(\alpha,\beta/\varphi_1,\varphi_2,|\psi\rangle)=\frac{1-\alpha\beta\cos{(\varphi_1-\varphi_2)}}{4}$ which is the quantum prediction for the singlet state. In later works Argaman, and Di Lorenzo \cite{Argaman2010,Lorenzo2012} rediscovered  the model with a more symmetric probability density $\rho(\theta_1,\theta_2/\varphi_1,\varphi_2,|\Psi_{EPR})=\frac{1}{4}\delta(\theta_2-\theta_1-\pi)[\delta(\theta_1-\varphi_1)+\delta(\theta_1-\varphi_1-\pi)+\delta(\theta_2-\varphi_2)+\delta(\theta_2-\varphi_2-\pi)]$.  We will come back to the interesting properties of these models concerning statistical independence and local-causality. Moreover for the moment the crucial issue is that these various $\theta$ variables have not to be interpreted as hidden or supplementary variables $\lambda$ (even if it was the interpretations made by the authors of these models). Instead, these models provide ontic representations of a complete quantum theory. In other words, examples provided by Beltrametti and Bugajski, or Scully and others explicitly show that Eq.~\ref{Bellmodel3} makes always sense even for approaches where QM is supposed to be complete. This is the reason why Eq.~\ref{Bellmodel3} can be seen as the natural generalization of Bell's formalism. \\ 
\indent In his later writings of 1976 and 1991 (\cite{Bell}, chaps. 7,24)  Bell  emphasized the importance of the generality of the beable concepts and made an explicit use of Eq.~\ref{Bellmodel3} for general beables $\omega$. In his famous article `Bertlmann's socks and the nature of reality' Bell wrote:
\begin{quote}
It is notable that in this argument  [i.e., Bell and EPR theorems] nothing is said about the locality, or even localizability, of the variable $\lambda$ [our $\omega$]. These variables could well include, for example, \textbf{quantum mechanical state vectors}, which have no particular localization in ordinary space-time. It is assumed only that the outputs A and B, and the particular inputs a and b [i.e., $\mathbf{\hat{n}}_1,\mathbf{\hat{n}}_2$], are well localized. \cite{Bell}, chap. 16, pp. 153-154.
\end{quote}    
In my opinion, beside the issue of locality this important point concerning $\omega$ notations answers some commentators who unfortunately still continue to believe that hidden variables are a prerequisite of Bell and EPR deductions. For example, in their 2014 QBist manifesto Fuchs, Mermin and Schack wrote: 
\begin{quote}
The parameter $\lambda$ [again our $\omega$] is undefined. It does not appear in the quantum theory. Nor has anybody ever suggested what in the experience of an agent $\lambda$ [$\omega$] might correspond to. In QBism this puts it outside the scope of physical science. \cite{Fuchs2014}
\end{quote} 
Clearly, the counterexamples of Beltrametti and Bugajski, or Scully illustrate the error. Furthermore, this proves that Bell's formalism used for discussing nonlocality of QM is actually independent of statements 	about realism or hidden variables. Therefore, this implies that repeated claims trying to oppose locality and realism as different alternatives to relinquish are badly motivated and based on misunderstanding of EPR and Bell works.   \\      
\indent Moreover, once we accept Eq.~\ref{Bellmodel3} we can go back to the EPR-Bell reasoning. This allows us to give  with Bell a rigorous definition of Einstein-locality or as Bell says  `local-causality'. Considering the elementary probability $dP(\alpha,\beta,\lambda,\theta/\mathbf{\hat{n}}_1,\mathbf{\hat{n}}_2,|\psi\rangle)$ we have always
\begin{eqnarray}
 dP(\alpha,\beta,\lambda,\theta/\mathbf{\hat{n}}_1,\mathbf{\hat{n}}_2,|\psi\rangle)=P(\alpha,\beta/\omega,\mathbf{\hat{n}}_1,\mathbf{\hat{n}}_2,|\psi\rangle)\rho(\omega,\mathbf{\hat{n}}_1,\mathbf{\hat{n}}_2,|\psi\rangle)d\omega \nonumber\\
 =P(\alpha/\beta,\omega,\mathbf{\hat{n}}_1,\mathbf{\hat{n}}_2,|\psi\rangle)P(\beta/\omega,\mathbf{\hat{n}}_1,\mathbf{\hat{n}}_2,|\psi\rangle)\rho(\omega,\mathbf{\hat{n}}_1,\mathbf{\hat{n}}_2,|\psi\rangle)d\omega.
 \end{eqnarray}
 Now, Bell showed through several important papers that the good definition of local causality implies actually three mathematically precise conditions:
 \begin{eqnarray}
 P(\alpha/\beta,\omega,\mathbf{\hat{n}}_1,\mathbf{\hat{n}}_2,|\psi\rangle)=P_1(\alpha/\omega,\mathbf{\hat{n}}_1,|\psi\rangle)\label{cond1}\\
P(\beta/\omega,\mathbf{\hat{n}}_1,\mathbf{\hat{n}}_2,|\psi\rangle)=P_2(\beta/\omega,\mathbf{\hat{n}}_2,|\psi\rangle)\label{cond2}\\
\rho(\omega,\mathbf{\hat{n}}_1,\mathbf{\hat{n}}_2,|\psi\rangle)=\rho(\omega,|\psi\rangle).\label{cond3}
 \end{eqnarray}
These conditions have been the subject  of intense debates in the past concerning locality and causality.  The last Eq.~\ref{cond3} is often named setting or measurement independence  in the literature (and sometimes freedom of choice condition) meaning that the distribution of beable $\omega$ prepared initially at the source is naturally expected in any `good' causal theory to be independent of the  directions $\mathbf{\hat{n}}_1,\mathbf{\hat{n}}_2$ that could be selected by macroscopic devices located in the remote past along the backward cones, e.g., by photons coming from far-away quasars existing billions of light years away of Alice and Bob. If we relax this condition we would get some fatalistic or superdeterministic theories (including for instance retrocausal models). Accepting Bell's condition Eq.~\ref{cond3} implies rejecting these fatalistic possibilities or loopholes (this natural assumption is accepted even in the nonlocal model of de Broglie and Bohm).  The two other conditions involve some hypotheses about outcome and parameter/setting independence. In particular, assuming again no superdeterminism and assuming  that the space-time points $x_1$, $x_2$ of the two recording events by Alice and Bob are space-like separated we obtain with Eqs.~\ref{cond1} and \ref{cond2} the more precise formulation of Einstein-locality quoted before forbidding  spooky and parasitic communications.\footnote{That was clearly the great achievement of Aspect group in the 1980's  to have experimentally developed such a configuration closing the communication loophole.}\\
\indent The following step for rederiving EPR  is now straightforward: by assuming Einstein/Bell locality, i.e., Eqs.~\ref{cond1}-\ref{cond3}, we deduce  for every  $\omega$ such that $\rho(\omega,|\psi\rangle)\neq 0$
\begin{eqnarray}
P_1(\alpha/\omega,\mathbf{\hat{n}},|\psi\rangle)P_2(\alpha/\omega,\mathbf{\hat{n}},|\psi\rangle)=0 \label{ded1}\\
P_1(\alpha/\omega,\mathbf{\hat{n}},|\psi\rangle)P_2(-\alpha/\omega,\mathbf{\hat{n}},|\psi\rangle)\geq 0\label{ded2}
\end{eqnarray}   
These conditions look harmless, but now suppose that for a given  actualized $\omega$ the state $|+1_{\mathbf{\hat{n}}}\rangle_1|-1_{\mathbf{\hat{n}}}\rangle_2$ was recorded as an outcome.  From Eq.~\ref{ded2}  it means that $P_1(+1/\omega,\mathbf{\hat{n}},|\psi\rangle)\neq 0$, $ P_2(-1/\omega,\mathbf{\hat{n}},|\psi\rangle)\neq 0$ and thus from Eq.~\ref{ded1} we must have $P_1(-1/\omega,\mathbf{\hat{n}},|\psi\rangle)=P_2(+1/\omega,\mathbf{\hat{n}},|\psi\rangle)=0$ and from probability conservation $P_1(+1/\omega,\mathbf{\hat{n}},|\psi\rangle)=P_2(-1/\omega,\mathbf{\hat{n}},|\psi\rangle)=1$. A similar situation would occur if for the given $\omega$ the result $|-1_{\mathbf{\hat{n}}}\rangle_1|+1_{\mathbf{\hat{n}}}\rangle_2$ obtained. In other words, we derive a deterministic theory: The EPR state requires  probabilities 
\begin{eqnarray}
P_1(\alpha/\omega,\mathbf{\hat{n}}_1,|\psi\rangle)=\delta_{\alpha,A(\lambda,\mathbf{\hat{n}}_1,|\psi\rangle)}=\frac{1+\alpha\cdot A(\lambda,\mathbf{\hat{n}}_1,|\psi\rangle)}{2}\label{finus}
\end{eqnarray} and similarly for $P_2(\beta/\omega,\mathbf{\hat{n}}_2,|\psi\rangle)$ with values zero or one and beables\footnote{Generally from  Eqs.~\ref{cond1}-\ref{cond3}  we can define local conditional mean value $\bar{A}(\lambda,\mathbf{\hat{n}}_1,|\psi\rangle)=\sum_{\alpha=\pm 1}\alpha P_1(\alpha/\omega,\mathbf{\hat{n}}_1,|\psi\rangle)$, and $\bar{B}(\lambda,\mathbf{\hat{n}}_2,|\psi\rangle)=\sum_{\beta=\pm 1}\beta P_2(\beta/\omega,\mathbf{\hat{n}}_2,|\psi\rangle)$  with $\sum_{\alpha=\pm 1} P_1(\alpha/\omega,\mathbf{\hat{n}}_1,|\psi\rangle)=\sum_{\beta=\pm 1}P_2(\beta/\omega,\mathbf{\hat{n}}_2,|\psi\rangle)=1$ and thus  $|\bar{A}(\lambda,\mathbf{\hat{n}}_1,|\psi\rangle)|\leq 1$, $|\bar{B}(\lambda,\mathbf{\hat{n}}_2,|\psi\rangle)|\leq 1$. Moreover, in the particular case of a deterministic theory  we have  $P_1(\alpha/\omega,\mathbf{\hat{n}}_1,|\psi\rangle)=1$ or $0$ and therefore $\bar{A}(\lambda,\mathbf{\hat{n}}_1,|\psi\rangle):=A(\lambda,\mathbf{\hat{n}}_1,|\psi\rangle)=\pm 1$ (and similarly for Bob side). Therefore, we deduce Eq.~\ref{finus}.}  $A(\lambda,\mathbf{\hat{n}},|\psi\rangle)=-B(\lambda,\mathbf{\hat{n}},|\psi\rangle)=\pm 1 $. From this derivation of determinism all the previous EPR deductions follow and in particular the logical results expressed by Eqs.~3,4. Most importantly the definition of QM-L and its negation QM-NL are transparent: This concludes our rederivation of EPR theorem. \\
\indent We stress that in the case QM-C using beables $\omega$ is a priori not mandatory to derive the EPR contradiction \cite{Norsen2006,Norsen2011}. Indeed, accepting locality is equivalent to assume the constraint 
$P(\alpha,\beta/\mathbf{\hat{n}}_1,\mathbf{\hat{n}}_2,|\psi\rangle)=P(\alpha/\mathbf{\hat{n}}_1|\psi\rangle)P(\beta/\mathbf{\hat{n}}_2|\psi\rangle)$ which is obviously wrong \cite{Norsen2006,Brukner2014}. Moreover, using $\omega$ leads to a more general and complete proof.
\section{The local realism rethoric}
\indent Before to conclude there is still one issue that we must discuss: the local realism controversy debated in several places. The local realism rhetoric originates from a very serious definition of `objective local theory' or `local realistic theory' by Clauser, Horne and Shimony in the 1970's as an honest substitute to the pejorative term `local hidden variables' \cite{Laudisa2023}. Moreover, the idea that one can actually decouple the realism from the locality in EPR and Bell's theorems is very strange and results from a misunderstanding.   Indeed, from the three alternatives listed in (4)  only (iii), i.e., ``QM-C $\&$ QM-NL'' refers to QM being complete. Bell's theorem rejecting alternative (i), we are left with (ii) ``QM-IC $\&$ QM-NL''  and (iii). But, note that the meaning of nonlocality changes from theory to theory. An advocate of complementarity will, following Bohr, more probably speaks about quantum wholeness, indivisibility of phenomena, or  non-separability in order to stress the difference with the action-at-a-distance semantic of Newtonian classical mechanics or hiddden variables \`a la dBB. Therefore, with such rephrasing the two surviving alternatives of Bell's theorem read:  
\begin{eqnarray}
(ii) \textrm{QM-NL}  & \& &  \textrm{QM-IC}\nonumber\\
(iii)' \textrm{QM-NS}  & \sim &  \textrm{QM-C}\label{EPR3b}
\end{eqnarray} where QS-NS is an abbreviation for quantum non-separable or anything similar and the `$\sim$' symbol is introduced to emphasize the sloppy equivalence or link between completness and Bohr's non-separability/indivisibility.  With this sloppy definition we have indeed to choose between nonlocal hidden  variables (ii) i.e., a particular form of realism  and between a version of QM  (iii)' considered as being complete and where the notion of  Einstein locality is analyzed  as too dogmatic and naive. The advocates of this rethoric will eventually rephrase (iii)' as just  quantum non realism or antirealism  (in relation with some form  of positivism, instrumentalism and/or operationalism) in order to contrast their view with the `naive' classical realism defended by Einstein or even Bohm. Therefore we end up with either assuming a realist nonlocal quantum world or a quantum nonrealist approach where `non separability' is the rule.\\
\indent However, in order to understand how  locality is (falsely) proposed as an alternative to realism it is crucial to see that the advocates of sloppy language often confuse the precise definition of local-causality (i.e.,  Eqs.~\ref{cond1}-\ref{cond3}, proposed by Bell after Einstein) with the also precise notion of local commutativity used in quantum-field and quantum measurement theory (for other related issues see also the Appendix). However, Bell was very clear (\cite{Bell}, chaps. 7,24) these two notions shouldn't be confused. Local commutativity $[\mathcal{O}_A,\mathcal{O}_B]=0$ of two local Hermitian operators $\mathcal{O}_A$ and $\mathcal{O}_B$ defined in two space-like separated spatial regions can be used to justify some averaged statistical independence of local measurements made by Alice and Bob. More precisely, consider for example the case where Alice is considering the evolution of the mean value $\langle\mathcal{O}_A(t)\rangle$ between times $t$ and $t+\delta t$ when Bob disturbs locally his settings. Assuming in the interaction picture  that $\mathcal{O}_A(t+\delta t)=\mathcal{O}_A(t)$ and that the quantum state evolves as $|\Psi(t+\delta t)\rangle=e^{-i\delta t\mathcal{O}_B(t)}|\Psi(t)\rangle$  we obtain if  $[\mathcal{O}_A,\mathcal{O}_B]=0$
\begin{eqnarray}
\langle\mathcal{O}_A(t+\delta t)\rangle= \langle \Psi(t)|e^{+i\delta t\mathcal{O}_B(t)}\mathcal{O}_A(t)e^{-i\delta t\mathcal{O}_B(t)}|\Psi(t)\rangle=\langle\mathcal{O}_A(t)\rangle.
\end{eqnarray} This condition shows that a local measurement made by Bob on his side cannot affect statistical observables of Alice. This is used to demonstrate the nonsignaling theorem which is a key feature of relativistic QM: $\sum_{\beta}P(\alpha,\beta/\mathbf{\hat{n}}_1,\mathbf{\hat{n}}_2,|\psi\rangle)=P(\alpha/\mathbf{\hat{n}}_1,\mathbf{\hat{n}}_2,|\psi\rangle)=P(\alpha/\mathbf{\hat{n}}_1,|\psi\rangle)$. Local-causality (i.e.,  Eqs.~\ref{cond1}-\ref{cond3} implies nonsignaling but the opposite is not always true. Still it is possible for a quantum nonrealist to use locality in the weak sense of nonsignaling and local commutativity and at the same time to consider quantum theory as requiring an indivisibility or non-separability of phenomena \`a la Bohr, i.e.,  what EPR and Bell define as (iii) ``QM-C $\&$ QM-NL''.  Clearly, it is the lack of precision in the language that allows something to be local and nonlocal at once. It is therefore not surprising that the false alternative between relinquishing locality or realism occurs if we use locality with such a wavy definition in the EPR and Bell theorems.  
\section{Conclusion: From EPR to Bohm and Bell}
\indent Going back to the EPR three alternatives (i), (ii) and (iii) of (4) we see that abandoning locality implies to relax at least  one of the three conditions \ref{cond1}-\ref{cond3}. Moreover, it is  remarkable that from the three possible alternatives of (4), only physical examples of case (ii) and (iii) are available.\\ 
\indent Consider first the case (iii) where QM is complete and nonlocal: We have already two representations given by Eq.~\ref{beltra} and \ref{scully}. In the representation of Beltrametti and Bugajski of Eq.~\ref{beltra} we see that Eq.~\ref{cond3} is preserved but we relinquish Eqs.~\ref{cond1}, \ref{cond2}. Using the representation of Scully et al. (i.e., Eq.~\ref{scully}) we now see that   Eqs.~\ref{cond1}, \ref{cond2} are preserved but actually it is Eq.~\ref{cond3} which is abandoned. This explains why this representation can be said to imply superdeterminism or retrocausality. Moreover, here the conditional probabilities  $P(\alpha/\theta_i,\varphi_i,|\psi\rangle)=\frac{1+\alpha\cos{(\varphi_i-\theta_i)}}{2}$ can only take value zero or one because of the specific form of $\rho(\theta_1,\theta_2/\varphi_1,\varphi_2,|\Psi_{EPR})$. The theory is thus `effectively' deterministic even though $P(\alpha/\theta_i,\varphi_i,|\psi\rangle)$ is in general different from  $0$ or $1$. This was expected because the derivation leading to  Eq.~\ref{finus} doesn't require Eq.~\ref{cond3} to hold true but only Eqs.~\ref{cond1}, \ref{cond2}. Furthermore, we see that once we add to Eq.~\ref{finus} the statistical independence given by Eq.~\ref{cond3} we obtain the EPR contradiction leading to the rejection of QM-C $\&$ QM-L theories.\\
\indent Now concerning case (ii): There is no known example of quantum theory involving hidden variables $\lambda$ that is at the same time local in Einstein-Bell sense. For instance, as it is well known the deterministic pilot wave theory of de Broglie and Bohm preserves the statistical independence of Eq.~\ref{cond3} but relax Eqs.~\ref{cond1}, \ref{cond2} in order to allow for action-at-a-distance with instantaneous forces. As Bohm wrote:
\begin{quote}
Thus the ``quantum-mechanical'' force may be said to transmit uncontrollable disturbances instantaneously from one particle to another through the medium of the $\psi-$field. \cite{Bohm1952}, p. 186.
\end{quote}
Bohm subsequently remarked that in his hidden variable theory, there is clearly a strong difference with the case of a single particle where the uncertainty principle explains why disturbance can locally preclude measurements of non-commuting observables within one single experimental protocol.  For entangled particles in the EPR case assuming hidden variables one should  perhaps  always expect a kind of nonlocal action-at-a distance affecting the two systems. This would somehow save the logics of the Heisenberg principle but would be in tension with special relativity.\\
\indent It was moreover Bell and not Bohm who answered the question: Is it possible to find an example of a quantum theory incomplete and local (i.e., case (i) of Eq.~8)? As we all know now the answer he provided was: No. It is not here the aim to review the derivation of Bell's theorem which only requires the validity of the 3 local causality conditions Eqs.~\ref{cond1}-\ref{cond3}. We point out that Bell's theorem is easily generalizable for stochastic hidden variables theories. This makes sense if the singlet state is interacting with non efficient detectors involving losses In this regime the EPR derivation leading to determinism is not generally valid but Bell's theorem and the contradiction with locality is preserved. The open and important question is of course which condition must be relaxed in order to develop a more complete theory involving hidden variables.\\
\indent  This problem will not be discussed here but just to mention that the choice followed by Bohm involves action-at-a-distance and a preferred space-time foliation conflicting with the goal of special and general relativity  which was to remove or prohibit such a preferred foliation  from physics (see   \cite{Drezet2019} for more on that issue and on how we can make sense of special foliations in the de dBB theory).  Moreover, in the approach of Bohm the mediation of such supra-luminal forces requires  a kind of `subquantum Aether' with remarkable properties. Indeed, Bohm's quantum potential acting between particles has several peculiar properties:  (i) Quantum potentials don't  decay with the distance (i.e., unlike gravitational Newtonian forces), (ii) quantum potentials are highly selective (e.g. entanglement and nonlocality can concerns two remote particles 1 and 2 but in turn ignore completely other `spectator' particles located near particles 1 and 2). Therefore, the quantum potential doesn't apparently spread in space unlike gravitational forces. (iii) Furthermore,  the quantum potential can correlate/entangle particles whatever the obstacles and barriers located between two particles. The quantum force is thus one of the most penetrative things of the Universe \cite{Elitzur1990}! (iv) At the same time, we never directly detected the presence of such a quantum potential (e.g., all attempts to directly detect `empty-waves' of such a quantum potential have for now failed). Therefore, the quantum potential seems not to carry energy by itself, i.e.,  it is not a field  existing independently of entangled particles and is apparently unseparable of the particles involved in the entangled wavefunction. Clearly, even in a Newtonian Universe, with instantaneous action-at-a-distance, such properties seem extroardinary or miraculous. Therefore, a subquantum Aether would be highly non-classical.  Moreover, even if one could understand and model the physical properties of such a subquantum Aether (attempts have been made for instance by Bohm and Vigier) this would probably be seen as a form of regression to pre-relativistic physics (because of a preferred foliation), and for many, including the present author, this would be too hard to swallow! Of course, this  doesn't unvalidate the dBB theory. Following Bell a number of prominent researchers have advocated a minimalistic position, i.e., accepting  the results of `Bohmian' mechanics in the configuration space and avoiding physical discussions about the physical nature or origin of the quantum potential and its remarkable nonlocal properties. With this minimalistic approach we can recover all the statistical predictions of standard quantum mechanics including of course the experimentally observed Bell's inequality violations.  However, if according to de Broglie, Bohm or Einstein the goal of physics is not only to predict but to explain then the minimalistic dBB theory is not a complete theory.\\       
\indent  For this reason, the present author progressively advocated an alternative `superdeterministic' approach \cite{Drezet2023} where the action-at-a-distance implied by the pilot-wave  of de Broglie and Bohm is preserved as an effective description.  At a lower `subquantum level' the theory is fully local in the sense that no signal can propagate faster than light but an underlying superdeterministic and time-symmetric fundamental field driving and synchronizing the particles allows us to relax  Eqs.~\ref{cond1}-\ref{cond3}  without contradicting the spirit of Einstein relativity. More precisely, in this new theory inspired by de Broglie `double-solution' work \cite{Broglie1925} particles are singularities of a classical field moving in the 4D space-time. The trajectories of the singularities obey the dBB  pilot-wave dynamics and the classical field is the half sum of retarded and advanced waves emitted by the singularities. The time-symmetry of the field allows us to derive nonlocality of the pilot-wave from an underlying superdeterministic and local theory.
\section{Appendix: Fine's theorem}
\indent A recurrent but misleading objection against the EPR-Bell deduction of nonlocality is based on a theorem attributed to A. Fine \cite{Fine1982} (already anticipated by G. Lochak in collaboration with de Broglie \cite{Lochak1976}). The claim (aldready debunked in \cite{Drezet2009}) is that assuming Bell's factorizability  for probability generally implies the existence of joint probabilities 
\begin{eqnarray}P(A_1,A_2,B_1,B_2)=\int P(A_1/\omega)P(A_2/\omega)P(B_1/\omega)P(B_2/\omega)\rho(\omega)d\omega \label{Fine}
\end{eqnarray} where $A_1$ and  $A_2$ (respectively $B_1$ and $B_2$) are two incompatible observables (i.e., spin components along the x and z directions) for particle 1 (respectively particle2).  Joint probabilities used in Bell's derivation  like $P(A_i,B_j)=\int P(A_i/\omega)P(B_j/\omega)\rho(\omega)d\omega$ are just marginals obtained from $P(A_1,A_2,B_1,B_2)$. Moreover, we can similarly obtain $P(A_1,A_2)=\int P(A_1/\omega)P(A_2/\omega)\rho(\omega)d\omega$ and since QM prohibits to define probability for incompatible/complementary observables for a same particle we have a clear contradiction. The claim of Lochak-Fine re-used, repeatedly by many authors (e.g., \cite{Brukner2014,Kupczynski2023} and references therein), is thus that Bell's reasoning implicitly takes for granted that such non-contextual  and non-physical joint probabilities like $P(A_1,A_2,B_1,B_2)$ exist. As written in \cite{Fine1982} `the existence of deterministic [Local] hidden variables violates the quantum mechanical condition that joint probability distributions are well defined only for commuting observables'. Again  (see for instance   \cite{Brukner2014}) the goal is to show that EPR-Bell local causality is a realist assumption which is dropped in QM.  However, this is misleading: nowhere in Bell's derivation using local causality it is mentioned that joint probability like $P(A_1,A_2,B_1,B_2)$ or $P(A_1,A_2)$ should exist (or if they exist they must not be interpreted as physical probabilities associated with measurements \cite{Drezet2009}). Nowhere it is assumed that the noncontextual factorizability used in Eq.~\ref{Fine} should be accepted. A formula like Eq.~\ref{Fine} can certainly be obtained for some noncontextual hidden-variables models but these models cannot agree with QM~\cite{Drezet2009}. 
         

\end{document}